\documentclass[aps,onecolumn,prd,nofootinbib]{revtex4}
\usepackage{amsmath}
\usepackage{graphicx}
\usepackage{dcolumn}
\usepackage{bm}
\usepackage{amssymb}
\usepackage{latexsym}
\usepackage{color}

\begin{document}

\title{Cosmic Duality and Statefinder Diagnosis of Spinor Quintom}

\author{Jing Wang$^{1,2,3}$\footnote{jwang@mx.nthu.edu.tw},~~Tian Lan$^{1}$,~~Tong-Jie Zhang$^{1,3}$}
\affiliation{$^1$Department of Astronomy, Beijing Normal University, Beijing, 100875, P.R. China
}

\affiliation{$^2$Institute of Astronomy, National Tsing Hua University,
Hsinchu 30013, Taiwan}

\affiliation{$^3$Center for High Energy Physics, Peking University, Beijing, 100871, P.R. China}

\begin{abstract}

In this paper, we study the possible
connections among different Spinor Quintom Dark Energy (DE) models
by the aid of duality. Then we apply the statefinder diagnostic to
these models. By this diagnostic pair {$\{r,s\}$}, we differentiate
one Quintom DE model from the others in a model independent manner.
A class of evolutionary trajectories of these Spinor Quintom models
are presented in the statefinder parameter planes. We also obtain
the current locations of the parameters $r$ and $s$, and these
locations correspond to different models in statefinder parameter
planes theoretically.

\end{abstract}

\keywords{Cosmology -- Observational cosmology -- Dark energy}

\maketitle

\section{Introduction}

There are mounting data from type Ia supernovae and cosmic microwave
background (CMB) radiation and so on \cite{1998snia, Spergel, Riess,
Seljak}. All these data have provided strong evidences for the
present universe, which is spatially flat, accelerated expanding and
dominated by dark sectors. The combined analysis of the above
cosmological observations supports that the contents of the universe
comprise about $73\%$ DE, $23\%$ cold dark matter (CDM), and only
$4\%$ usual baryon matter which can be described by the well-known
particle theory. In terms of Friedmann-Robertson-Walker (FRW)
cosmology, this acceleration is attributed to an exotic form with
negative pressure, the so-called DE. So far, the nature of DE
remains a mystery. Theoretically, the obvious candidate for such a
component is a small cosmological constant $\Lambda$ (or vacuum
energy) with Equation of State (EoS) $w=-1$. The corresponding
cosmological model
--- LCDM (or $\Lambda$CDM) --- consists of a mixture of vacuum
energy and CDM, but there exsits the fine tuning and the coincidence
problems. The inspiration of inflation suggests that DE is
attributed to the dynamics of a scalar field or multi-scalar fields,
such as the Quintessence \cite{Wetterich:1987fm, Ratra:1987rm},
Phantom \cite{Caldwell:1999ew}, K-essence
\cite{ArmendarizPicon:2000ah, Chiba:1999ka, Caldwell:2003vq}. There
are also other DE models such as Chaplygin gas
\cite{Kamenshchik:2001cp}, Braneworld models \cite{Dvali:2000hr,
Deffayet:2001pu}, Holographic models \cite{Fischler:1998st,
Cohen:1998zx, Hsu:2004ri, Li:2004rb}, and so on. Although the recent
fits to the data from combination of WMAP \cite{Spergel:2006hy,
Komatsu:2008hk}, the recently released 182 SNIa Gold sample
\cite{Riess:2006fw} and also other cosmological observational data
show remarkably consistence of the cosmological constant, it is
worth noting that a class of dynamical models with EoS across $-1$
{\it Quintom} is mildly favored \cite{Feng:2004ad, Zhao:2006qg,
Zhao:2006bt, Wang:2006ts}. In the literature there have been a lot
on theoretical studies of Quintom-like models \cite{Cai:2007gs,
Xia:2007km, Li:2005fm, ArmendarizPicon:2004pm, Koivisto:2008xf,
Cai:2005ie, Aref'eva:2005fu, Guo:2004fq, Quintom_tf, Cai:2006dm,
Quintom_1, Cai:2007qw, Cai:2007zv, Cai:2008ed, Quintom_others,
Xiong:2007cn, Koivisto:2007bp, Elizalde:2004mq}.

Previously, it has been proved that a Quintom DE model and its
combination with Chaplygin gas fluid can be realized by non-regular
spinor matter \cite{Cai:2008gk}. Interestingly, this type of model
can realize many kinds of Quintom scenarios by transforming the form
of potential of the spinor. Two typical Quintom models among these
are Quintom-A and Quintom-B: the former with EoS $w > 1$ at early
time and $w < 1$ lately; while the latter with the EoS arranged and
changing from below $-1$ to above $-1$. To understand the possible
combinations among different types of Quintom model in spinor field,
we perform the implications of cosmic duality in this class of
models in this paper. Cosmic duality is a mathematic feature which
originates from string cosmology \cite{Veneziano:1991ek,
Lidsey:1999mc}. Later on, cosmic duality is used to connect the
standard cosmology with phantom cosmology and generalized into
studies in more complicated DE models (see Refs.
\cite{Chimento:2003qy, Dabrowski:2003jm, Chimento:2005xa,
Dabrowski:2006dd}). Ref. \cite{Cai:2006dm} was pointed out that
there is a dual behavior between the models of Quintom-A and
Quintom-B. By studying the behavior of the energy density and
pressure in spinor field, we find a duality between the Quintom-A
and Quintom-B. Meanwhile, we realize other Quintom models by
considering this property.

Since more and more DE models have been developed to explain the
current cosmic acceleration, a method for discriminating contenders
in a model independent manner was proposed by Sahni in Ref.
\cite{Chiba:1998tc, Sahni:2002fz}. The new cosmological diagnostic
pair $\{r,s\}$, called statefinder, is a geometrical diagnostics.
This diagnostics is algebraically related to the higher derivatives
of the scale factor $a$ with respect to time. It seems a natural
next step to study beyond the Hubble parameter
$H\equiv\frac{\dot{a}}{a}$ and the deceleration parameter $q$.
Because the model-dependent physical variables describing the DE
depend on the properties of physical fields, the DE models can be
distinguished more effectively by statefinder than these variables.
Hitherto, some DE models have been perfectly differentiated, such as
LCDM universe, Quintessence, Phantom, the Chaplygin gas, Braneworld
models, Holographic models, and interacting and coupling DE models.
Correlative work has been performed by Ref. \cite{Zimdahl:2003wg,
Alam:2003sc, Gorini:2002kf, Zhang:2005rj, Zhang:2005yz, Wu:2005apa,
Zhang:2004gc, Setare:2006xu, Yi:2007gu, Zhao:2007jn,
Panotopoulos:2007zn, Huang:2008gs, Shao:2008zz, Feng:2008rs}. We
apply diagnostic to the Spinor Quintom models and present the
trajectories in the $r-s$ plane corresponding to these kinds of
Quintom DE models. The fixed point $\{r,s\}=\{1,0\}$ is in
correspondence with the spatially flat LCDM scenario. For one given
Quintom model in spinor field, the departure from the fixed point
$\{r,s\}=\{1,0\}$ provides a nice way to determine the distance from
LCDM.

This paper is organized as follows:
In section 2, we realize possible connections among the different
spinor Quintom DE models using the cosmic duality. In section 3, we
apply the statefinder diagnostic to these Spinor Quintom DE models.
Section 4 gives discussions and conclusions.

\section{Duality of Spinor Quintom Universes}

In this section, we investigate the
possible connections among different Spinor Quintom models and
realize more Quintom DE models by the aid of the cosmic duality. The
cosmic duality has been investigated by many work. The authors of
Ref.\cite{Chimento:2002gb,
Aguirregabiria:2003uh,Aguirregabiria:2004te} have considered a
possible transformation with the Hubble parameter and studied the
relevant issues with the cosmic duality\cite{Veneziano:1991ek,
Lidsey:1999mc}. Specifically, Ref.\cite{Chimento:2003qy} has shown a
connection between a quintessence cosmology and a contracting
phantom cosmology. Later on, this duality has been generalized into
studies with more complicated DE models\cite{Dabrowski:2003jm,
Chimento:2005xa, Chimento4, Chimento:2006gk}. Moveover, The form
invariance transformations are used to constructed phantom cosmology
and extended to the fermion fields \cite{Chimento:2007fx}. Motivated
by these work, we investigate the cosmic duality in the Spinor
Quintom scenario under the transformation performed in Ref.
\cite{Chimento:2002gb, Aguirregabiria:2003uh,
Aguirregabiria:2004te}.

We consider a universe filled with Quintom DE perfect fluid in
spinor field \cite{ArmendarizPicon:2003qk, Vakili:2005ya,
Ribas:2005vr}, neglecting the contributions to the components of
matter and radiation. We deal with the homogeneous and isotropic FRW
space-time, and assume the space-time metric as,
\begin{equation}
ds^{2}=dt^{2}-a^{2}(t)d\vec{x}^2~.
\end{equation}
According to the dynamics of a spinor field which is minimally
coupled to Einstein's gravity\cite{Weinberg,BirrellDavies,GSW}, we
can write down the following Dirac action in a curved background
space-time
\begin{eqnarray}\label{action}
S_{\psi}&=&\int d^4 x~e~[\frac{i}{2}(\bar\psi\Gamma^{\mu}D_{\mu}
\psi-D_{\mu}\bar\psi\Gamma^{\mu}\psi)-V]\nonumber\\
&=&\int d^4 x ~e~{\cal L}_{\psi}~,
\end{eqnarray}
where, $e$ is the determinant of the vierbein $e_{\mu}^{a}$ and $V$
stands for any scalar function of $\psi$, $\bar\psi$ and possibly
additional matter fields. We assume that $V$ only depends on the
scalar bilinear $\bar\psi\psi$. For a gauge-transformed homogeneous
and a space-independent spinor field, the equation of motion for
spinor reads,
\begin{eqnarray}\label{EoM}
\dot{\psi}+\frac{3}{2}H\psi+i\gamma^{0} V' \psi&=&0~,\\
\dot{\bar\psi}+\frac{3}{2}H\bar\psi-i\gamma^{0}V' \bar\psi&=&0~,
\end{eqnarray}
where a dot denotes a time derivative and a prime is a derivative
with respect to $\bar\psi\psi$, while H is Hubble parameter.
Taking a further derivative, we can obtain the solution of
equation of motion:
\begin{equation}\label{solution}
\bar\psi\psi=\frac{N}{a^{3}}~,
\end{equation}
where $N$ is a positive time-independent constant and we define it
as the present value of $\bar\psi\psi$. From the expression of the
energy-momentum tensor in Ref. \cite{Cai:2008gk} and the equation of
motion for spinor, we get the energy density and the pressure of the
spinor field:
\begin{eqnarray}
\label{density}\rho_{\psi}&=&T_{0}^{0}=V~,\\
\label{pressure}p_{\psi}&=&-T_{i}^{i}=V'\bar\psi\psi-V~.
\end{eqnarray}
The EoS of the spinor field, defined as the ratio of its pressure
to energy density, is given by
\begin{equation}\label{eos}
w_{\psi}\equiv\frac{p_{\psi}}{\rho_{\psi}}=-1+\frac{V'\bar\psi\psi}{V}~.
\end{equation}
To keep the energy density positive, one may see that there are
$w_{\psi}>-1$ when $V'>0$ and $w_{\psi}<-1$ when $V'<0$ from Eq.
(\ref{eos}). The former corresponds to a Quintessence-like phase and
the latter stands for a Phantom-like phase. Therefore it requires
that the derivative of the potential $V'$ change its sign if one
expects a Quintom picture. In terms of the variations of $V'$, (1).
if $V'>0~~~\rightarrow~~~V'<0 $, we get a Quintom-A scenario which
evolves from a Quintessence-like phase with $w_{\psi} > -1$ to a
Phantom-like phase with $w_{\psi} < -1$; (2). while
$V'<0~~~\rightarrow~~~V'>0$, one can obtain a Quintom-B scenario,
and for which the EoS is arranged to and changes from below $-1$ to
above $-1$.


In the framework of FRW cosmology, the Friedmann equation reads,
\begin{equation}\label{FE}
H^2=\frac{1}{3}\rho~,
\end{equation}
where we use units $8\pi G=\hbar=c=1$ and all parameters are
normalized by $M_p=1/\sqrt{8 \pi G}$ in the paper.

The form-invariant transformation
\cite{Chimento:2002gb, Aguirregabiria:2003uh, Aguirregabiria:2004te}
reads,
\begin{eqnarray}
\bar\rho&=&\bar\rho(\rho)~,\\
\bar H&=&-(\frac{\bar\rho}{\rho})^\frac{1}{2}H~.
\end{eqnarray}
Then the corresponding changes for the pressure $p$ and the EoS $w$
are,
\begin{eqnarray}
\label{bpres}\bar p&=&-\bar\rho-(\frac{\bar\rho}{\rho})^\frac{1}{2}(\rho+p)\frac{d\bar\rho}{d\rho}~,\\
\label{bEoS}\bar
w&=&-1-(\frac{\bar\rho}{\rho})^\frac{3}{2}\frac{d\bar\rho}{d\rho}(1+w).
\end{eqnarray}

Taking $\bar\rho=\rho$ in Eqs. (\ref{bpres}) and (\ref{bEoS}) as an
example of detailed discussion, regardless of loss of the generality
of the physical conclusion and information, we can get the dual
transformation:
\begin{eqnarray}
\label{bH}\bar H&=&-H,\\
\label{bp}\bar p&=&-2\rho-p=-V'\bar{\psi}\psi-V,\\
\label{bw}\bar w&=&-2-w=-1-\frac{V'\bar{\psi}\psi}{V}.
\end{eqnarray}
Consequently, the dual form of Lagrangian reads,
\begin{equation}\label{lagrangian}
\bar{\cal L}=-\frac{i}{2}(\bar\psi\Gamma^\mu D_\mu\psi -
D_\mu\bar\psi\Gamma^\mu\psi)-V.
\end{equation}
Contrasting the Lagrangian derived from Eq. (\ref{action}) and its
dual form in (\ref{lagrangian}), we find that if the original
Lagrangian is for a Quintom-A model, and the dual one is for a
Quintom-B one by the dual transformation, and vice versa. With this
property, it is possible that one model for the evolution of the
universe can be connected to others. While from the Eq. (\ref{bw}),
one can expect a symmetrical
evolution tracks of the EoS comparing with its dual Eq. (\ref{eos}):\\
(i). There is
\begin{eqnarray}
V'<0~~~\rightarrow~~~V'>0 \nonumber~,
\end{eqnarray}
which gives a Quintom-A scenario by describing the universe, and the
universe evolves from Quintessence-like phase with $w_{\psi} > -1$
to
Phantom-like phase with $w_{\psi} < -1$; \\
(ii). There is
\begin{eqnarray}
V'>0~~~\rightarrow~~~V'<0 \nonumber~,
\end{eqnarray}
which gives a Quintom-B scenario and for this scenario, the EoS is
arranged and changes from below $-1$ to above $-1$; \\
Solving the Einstein equation (\ref{FE}) and its dual form
(\ref{lagrangian}), we may discuss different periods of the
evolution of Quintom universe.

To discuss in detail, we will consider three special kinds of
power-law-like potentials and perform its semi-analytic solution to
present the dual characteristics, then we take these potentials to
study its numerical solutions.

\begin{figure}[htbp]
\includegraphics[scale=0.8]{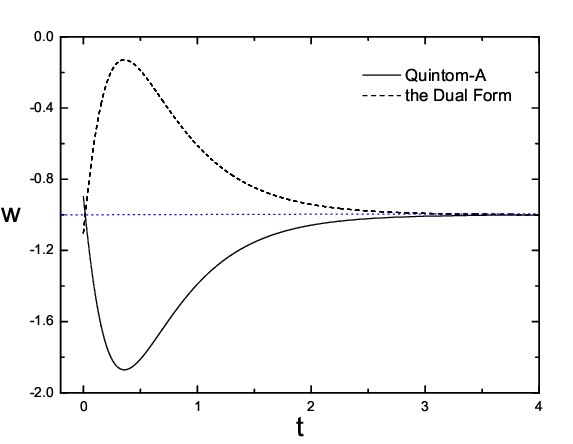}
\caption{The evolution of Quintom-A and its dual Quintom-B as a
function of time for $V=V_0[(\bar\psi\psi-b)^2+c]$, where
$V_0=1.0909\times10^{-117}$. The model parameters $b=0.05,
c=10^{-3}$. The initial conditions $(\bar\psi\psi)_0=0.051$.
\label{Fig:fig1}}
\end{figure}

In the first instance, we take a potential as
$V=V_0[(\bar\psi\psi-b)^2+c]$ which can realize Quintom-A scenario,
and its detailed discussion can be found in Ref. \cite{Cai:2008gk}.
Its dual solution is a description of the universe in the case of
Quintom-B. According to Eq. (\ref{solution}), one finds that
$\bar\psi\psi$ is decreasing along with an increasing scale factor
$a$ during the expansion of the universe. From the formula of $V'$,
we deduce that at the beginning of the evolution the scale factor
$a$ is very small, and so correspondingly $\bar\psi\psi$ becomes
very large in order to ensure $V'>0$ at the beginning. Then
$\bar\psi\psi$ decreases along with the expanding of $a$. At the
moment of $\bar\psi\psi=b$, one can see that $V'=0$ which results in
the EoS $w_{\psi}=-1$. After that $V'$ becomes less than $0$, so
then the universe enters a Phantom-like phase. Finally, the universe
approaches to a de-Sitter space-time in the Quintessence Phase in
the future. Accordingly, the EoS of the dual form evolves from below
$-1$ and crosses $-1$ as $t\rightarrow 0$, later, to above $-1$. In
final, it approaches to the cosmological constant when $t\rightarrow
+\infty$. It is shown that either Quintom-A or Quintom-B will avoid
a big rip when $w<-1$. In Fig. \ref{Fig:fig1}, we plot the concrete
picture of this dual pair. We can find that the evolution of these
two models is just as a dual process fit.

\begin{figure}[htbp]
\begin{center}
\includegraphics[scale=0.8]{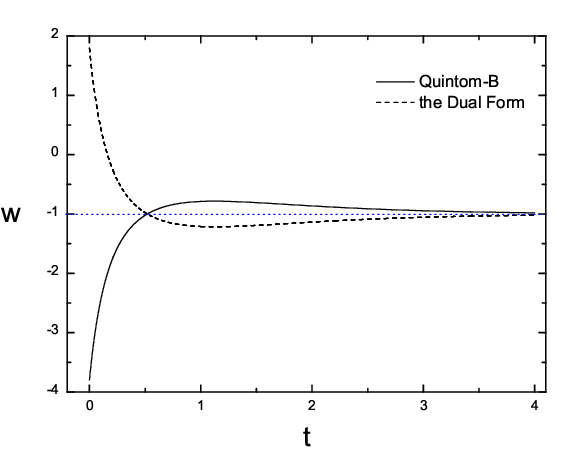}
\caption{Same as Fig.1, but for the evolution of Quintom-B and its
dual Quintom-A. The solid line is the Quintom-B model, and the dash
line is its dual model. \label{Fig:fig2}}
\end{center}
\end{figure}

In succession, if $V=V_0[-(\bar\psi\psi-b)\bar\psi\psi+c]$, we
obtain a Quintom-B model (see Ref. \cite{Cai:2008gk}). Taking a dual
form theoretically as discuss above, we can present a numerical
solution in Fig. \ref{Fig:fig2}. Clearly, the duality of this Spinor
Quintom model shows a the evolutionary picture of Quintom-A.

These two kinds of models describe different behaviors of the
cosmological evolution: one is an expanding phase while the other
lies in the contracting one. The behavior depends on the potential
and initial conditions we choose. It is found that Quintom model and
its dual form are symmetrical with respect to $w=-1$.

\begin{figure}[htbp]
\begin{center}
\includegraphics[scale=0.8]{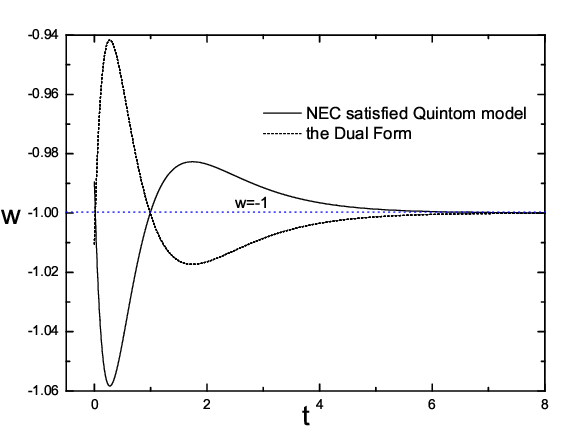}
\caption{Same as Fig.1, but for the evolution of NEC satisfied
Quintom model and its dual model. The solid line is the Quintom-B
model, and the dash line is its dual model. \label{Fig:fig3}}
\end{center}
\end{figure}

For an extended investigation, we take
$V=V_{0}[(\bar\psi\psi-b)^{2}\bar\psi\psi+c]$, which can realize a
picture across $w=-1$ twice. In Fig. \ref{Fig:fig3}, we can see that
this dual model evolves from below $-1$ and crosses $-1$ twice, and
later to below $-1$ again. Ultimately, it approaches to the
cosmological constant boundary when $t\rightarrow+\infty$.

From the above analysis, we investigate the connections among
different models and evolutionary trajectories of our universe with
the help of this characteristic. It is known that under different
kinds of model of evolution, the fate of the universe will be
different. Our study in this section helps us understand the
properties of various DE models and their connections to the
evolution and the fate of the Universe. Moreover, the past and
future properties can be understood by studying the above
characteristics. One application to combine these properties
together is Quintom-like bouncing cosmology, which has been
intensively studied in Refs. \cite{Cai:2007qw, Cai:2007zv,
Cai:2008ed}. In the meantime, we also realize three more Quitom
models through the dual transformation.

\section{Statefinder diagnostic to Spinor Quintom models}

Based on the above discussions, we have known the connections
between two kinds of Spinor Quintom models. It can be seen that
there are so many Spinor Quintom DE models proposed to explain the
cosmic acceleration, thus how to distinguish these models become a
widely attentional issue. It is no
doubt that the effective EoS is an important property of DE, and
different model is described by different EoS. However, the EoS,
which depends on the selection of FRW background and the
pre-supposition for ignorance of the contributions to the components
of matter and radiation, is a model-dependent
parameter\cite{Sahni:2002fz}. In order to differentiate the
different DE models fundamentally, we need a class of variables
which reflect the fundamental property for field-theoretical DE
models and should be independent of any assumption when the models
are constructed. Considering this, we use one pair of widely used
parameter---statefinder parameter---to differentiate them in this
section. Also on the grounds of this consideration, Sahni proposed
the geometrical--constructed from space-time metric directly--
statefinder diagnostic pair $\{r,s\}$, which is defined as
\cite{Chiba:1998tc, Sahni:2002fz},
\begin{equation}
r\equiv\frac{\dddot{a}}{aH^{3}},~~~
s=\frac{r-1}{3(q-\frac{1}{2})},
\end{equation}
where $q$ is the deceleration parameter,
\begin{equation}
q=-\frac{\ddot{a}}{aH^{2}}.
\end{equation}
Accordingly, by showing different evolutionary trajectories
qualitatively in the $r-s$ and $r-q$ planes, this statefinder pair
can differentiate one DE model from the others.  In what follows we
will apply the statefinder diagnostic to three Quintom models in
spinor field---Quintom-A, Quintom-B, and the Quintom model crossing
$-1$ twice. We use the form of the statefinder parameter written by
pressure and energy density in the following text,
\begin{equation}
r=1+\frac{9(\rho+p)\dot{p}}{2\rho\dot{\rho}},~~s=\frac{(\rho+p)\dot{p}}{\rho\dot{\rho}},
\end{equation}
where the energy density and pressure are given by Ref.
\cite{Cai:2008gk}.

Taking components of dark matter and DE into account in a
spatially flat universe, we can write down the Friedmann equation:
\begin{equation}
H^{2}=\frac{1}{3}(\rho_\psi+\rho_m),
\end{equation}
where $\rho_m$ is the energy density of dark matter with EoS
$p_m=(\gamma_m-1)\rho_m$. Ignoring the interaction between the two
dark sectors, we can see that the energy density of both DE and dark
matter are conserved and satisfy its continuity equation,
respectively,
\begin{equation}
\dot{\rho_\psi}+3H(\rho_\psi+p_\psi)=0,
\end{equation}
\begin{equation}
\dot{\rho_m}+3H\rho_m=0.
\end{equation}
As a result, using the EoS, the equation of motion and the Friedmann
equation, we obtain the following expressions,
\begin{eqnarray}
r&=&1+\frac{9}{2}w_\psi(1+w_\psi)\Omega_\psi+\frac{9}{2}w'\overline{\psi}\psi\Omega_\psi~,\\
s&=&1+w_\psi+\frac{w'}{w_\psi}\overline{\psi}\psi~,
\end{eqnarray}
and the deceleration parameter
\begin{equation}
q=\frac{1}{2}+\frac{3}{2}w_\psi\Omega_\psi,
\end{equation}
where $\Omega_\psi=\frac{\rho_\psi}{\rho}$ is the fraction of energy
density of DE, and $w'$ is the derivative of EoS with respect to
$\overline{\psi}\psi$.

\begin{figure}[tbp]
\includegraphics[width=0.4\textwidth]{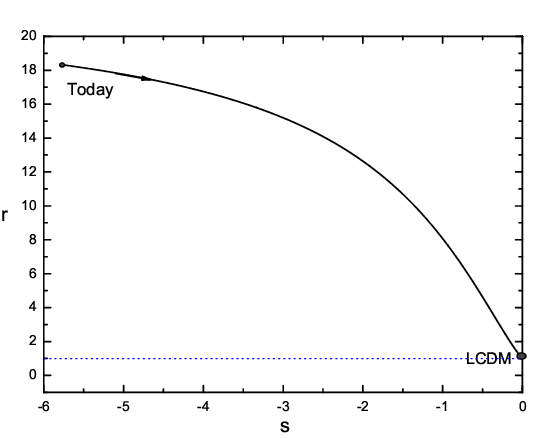}%
\includegraphics[width=0.4\textwidth]{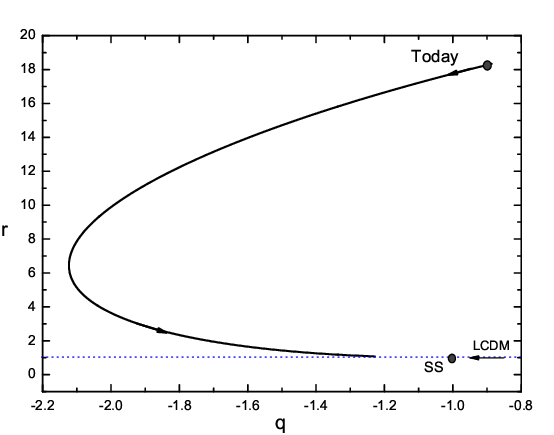}
\caption{The $r(s)$ and $r(q)$ for the potential
$V=V_0[(\bar\psi\psi-b)^2+c]$, where $V_0=1.0909\times10^{-117}$.
The model parameters $b=0.05, c=10^{-3}$. The initial conditions
$(\bar\psi\psi)_0=0.051$. \label{Fig:fig4}}
\end{figure}

Hereinafter, we study the statefinder diagnostic for the Spinor
Quintom models in three different potentials. Firstly, we discuss
the Quintom-A model with the form of potential
$V=V_0[(\bar\psi\psi-b)^2+c]$, where $V_0$, $b$, $c$ are undefined
parameters. In Fig. \ref{Fig:fig4}, we show the time evolution of
statefinder pair $\{r,s\}$ and $\{r,q\}$. In the left figure of Fig.
\ref{Fig:fig4}, the LCDM scenario corresponds to a fixed point
$s=0,~r=1$. It can be found that $r$ monotonically decreases with
$s$ from today to the point of LCDM along the trajectory of Spinor
Quintom-A model in $r-s$ plane. The current value also can be
presented.

\begin{figure}[tbp]
\includegraphics[scale=0.8]{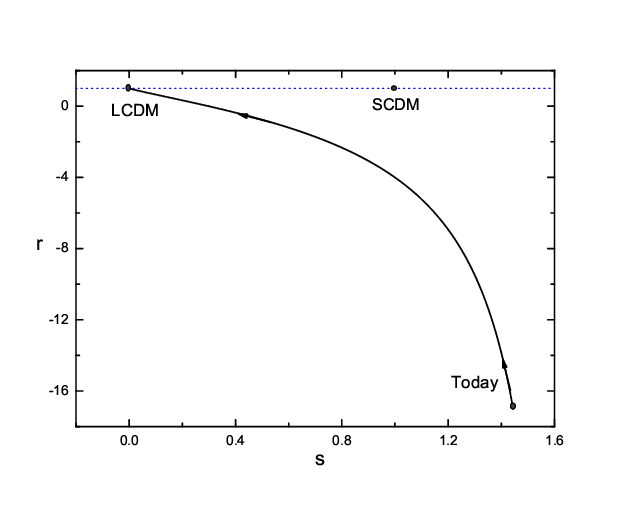}%
\includegraphics[scale=0.8]{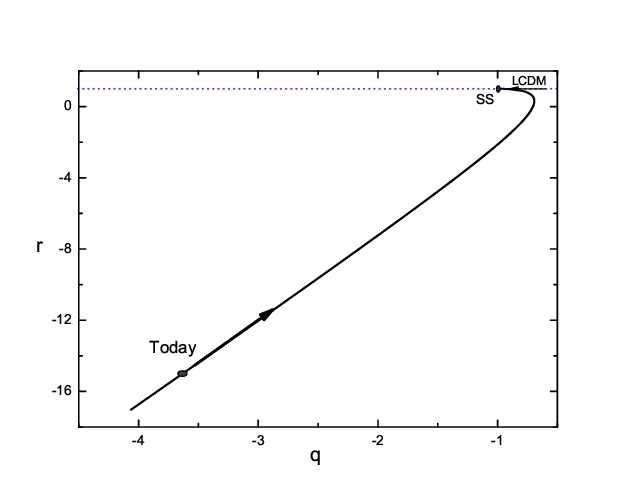}
\caption{Same as Fig.4 but for the potential
$V=V_0[-(\bar\psi\psi-b)\bar\psi\psi+c]$. The location of these two
tracks are roughly opposite to that of Fig. 4. \label{Fig:fig5}}
\end{figure}

Next, the trajectories of Spinor Quintom-B model with potential
$V=V_0[-(\bar\psi\psi-b)\bar\psi\psi+c]$ are plotted. In numerical
calculations, we take the same value as Quintom-A model. It can be
seen that the evolutionary graphics of Quintom-B is roughly opposite
to that of Quintom-A in both $\{r,s\}$ and $\{r,q\}$ planes (See
Fig. \ref{Fig:fig5} for a clear image). On the contrary, $r$
monotonically increases with $s$ from today to the point of $(0,1)$
corresponding to LCDM. Furthermore, this trajectory is always below
that of the standard cold dark matter(SCDM).

\begin{figure}[tbp]
\includegraphics[width=0.4\textwidth]{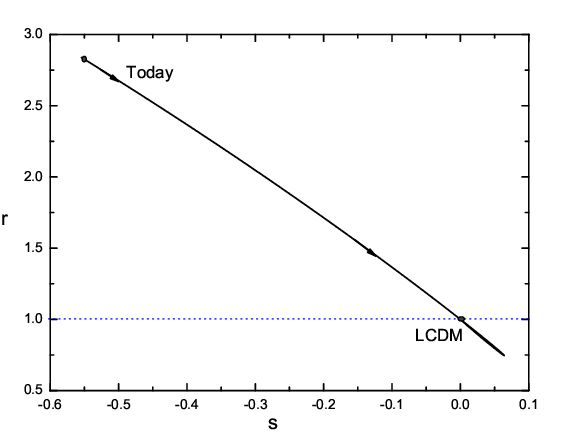}%
\includegraphics[width=0.4\textwidth]{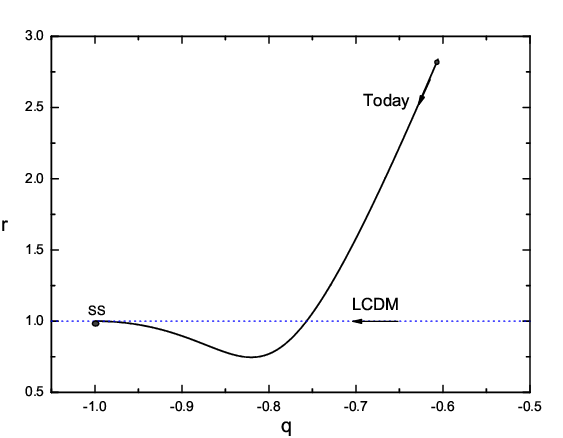}
\caption{Same as Fig.4 but for the potential
$V=V_0[(\bar\psi\psi-b)^2\bar\psi\psi+c]$.\label{Fig:fig6}}
\end{figure}

Finally, we turn to the case of crossing the cosmological boundary
twice. The phase portraits of $\{r,s\}$ and $\{r,q\}$ are presented
in Fig. \ref{Fig:fig6}, respectively, where the values of parameters
are also the same as those of the case of Quintom-A. In $r-s$ plane,
we can see that $r$ also monotonically decreases with $s$ from today
and then crosses the point of LCDM. And in final, it ends at this
point. While for $r-q$ diagram, both our model and LCDM evolve to a
steady state cosmology(SS) at when $q=-1,~r=1$, i.e. the de Sitter
expansion.

We can see that in $r-s$ diagram both of the two former models end
their evolution at the fixed point $\{r,s\}=\{1,0\}$ corresponding
to the LCDM. In the case of Quintom-A, it is found that the $r-s$
phase portrait is always in the region of negative $s$ and positive
$r$. On the contrary, most of the $r(s)$ trajectories of Quintom-B
lies in the opposite locations. While the third case gives another
evolutionary portraits. From the above numerical analysis, we find
that either of these Quintom model ends their evolution at a fixed
point---LCDM, in according with a De Sitter space-time realized in
the second section.

In conclusion, we have investigated the dynamics of Spinor Quintom
DE models by using the new geometrical diagnostic
method---statefinder pair $\{r,s\}$. The Ref. \cite{Wu:2005apa} has
applied this method to Quintom model and successfully differentiate
this class of models from other DE models. But it seems not useful
to discriminate Quintom models in different potentials. However, as
can be seen in this section, the statefinder diagnostic is able to
differentiate different Quintom models from diverse kinds of
power-law potential in the spinor scenario, as well as distinguish
the Spinor Quintom models from other DE models.
Current data is not precise enough
to distinguish these DE models by the aid of the statefinder pairs,
and we expect further and more exact data to constrain the
properties of DE.

\section{Conclusion and Discussions}

To summarize, since more and more Quintom DE models are proposed, we
established the connections among these models by the aid of the
cosmic duality. To connect the two totally different scenarios of
universe evolution the cosmic duality in spinor scenario, we keep
the energy density of the Universe and Einstein equations unchanged,
but transforming the Hubble parameter.
Besides, in order to make
fundamental and model-independent differentiation among different DE
models, we apply the new geometrical diagnostic method---statefinder
pair $\{r,s\}$ to the Spinor Quintom model. We differentiate
different Quintom models with different kinds of power-law
potentials in the spinor scenario, and distinguish the Spinor
Quintom models from other DE models, as well.
As we know, the statefinder
parameters $r-s$, which are the natural next step beyond the most
well-known and widely used geometrical parameters Hubble parameter
H(t) and the deceleration parameter q(t), relate to the third order
derivatives of scalar factor. It is enough to constrain the DE
models at present by the aid of this pair of parameters. However,
Current data is not precise enough to distinguish these DE models,
and we expect further and more exact data and the other parameters
to constrain the properties of DE.

\section*{Acknowledgements}

We are very grateful to the anonymous referee for many valuable
comments that greatly improved the paper. It is a pleasure to thank
Yi-Fu Cai, Tao-Tao Qiu and Xinmin Zhang for enlightening discussions
and cooperations at the beginning. We also thank Chao-Jun Feng for
helpful discussions. This work was supported by the National Science
Foundation of China (Grants No. 11173006), the Ministry
of Science and Technology National Basic Science program
(project 973) under grant No. 2012CB821804, and the Fundamental
Research Funds for the Central Universities.


\vfill

\begin{thebibliography}{99}

\bibitem{1998snia}
  S. Perlmutter {\it et al.},
  Astrophys. J. {\bf 483}, 565 (1997);
  Adam G. Riess et al.,
  Astrophys. J. {\bf116}, 1009 (1998).

\bibitem{Spergel}
  D. N. Spergel {\it et al.},
  Astrophys. J. Suppl. {\bf 148}, 175 (2003).

\bibitem{Riess}
  A. G. Riess {\it et al.},
  Astrophys. J. {\bf 607}, 665 (2004).

\bibitem{Seljak}
  U.~Seljak {\it et al.},
  Phys.\ Rev.\  D {\bf 71}, 103515 (2005)
  [arXiv:astro-ph/0407372].


\bibitem{Wetterich:1987fm}
  C.~Wetterich,
  Nucl.\ Phys.\  B {\bf 302}, 668 (1988).

\bibitem{Ratra:1987rm}
  B.~Ratra and P.~J.~E.~Peebles,
  Phys.\ Rev.\  D {\bf 37}, 3406 (1988).


\bibitem{Caldwell:1999ew}
  R.~R.~Caldwell,
  Phys.\ Lett.\  B {\bf 545}, 23 (2002)
  [arXiv:astro-ph/9908168].


\bibitem{ArmendarizPicon:2000ah}
  C.~Armendariz-Picon, V.~F.~Mukhanov and P.~J.~Steinhardt,
  Phys.\ Rev.\  D {\bf 63}, 103510 (2001)
  [arXiv:astro-ph/0006373].

\bibitem{Chiba:1999ka}
  T.~Chiba, T.~Okabe and M.~Yamaguchi,
  Phys.\ Rev.\  D {\bf 62}, 023511 (2000)
  [arXiv:astro-ph/9912463].

\bibitem{Caldwell:2003vq}
  R.~R.~Caldwell, M.~Kamionkowski and N.~N.~Weinberg,
  Phys.\ Rev.\ Lett.\  {\bf 91}, 071301 (2003)
  [arXiv:astro-ph/0302506].

\bibitem{Kamenshchik:2001cp}
  A.~Y.~Kamenshchik, U.~Moschella and V.~Pasquier,
  Phys.\ Lett.\  B {\bf 511}, 265 (2001)
  [arXiv:gr-qc/0103004].

\bibitem{Dvali:2000hr}
  G.~R.~Dvali, G.~Gabadadze and M.~Porrati,
  Phys.\ Lett.\  B {\bf 485}, 208 (2000)
  [arXiv:hep-th/0005016].

\bibitem{Deffayet:2001pu}
  C.~Deffayet, G.~R.~Dvali and G.~Gabadadze,
  Phys.\ Rev.\  D {\bf 65}, 044023 (2002)
  [arXiv:astro-ph/0105068].

\bibitem{Fischler:1998st}
  W.~Fischler and L.~Susskind,
  arXiv:hep-th/9806039.

\bibitem{Cohen:1998zx}
  A.~G.~Cohen, D.~B.~Kaplan and A.~E.~Nelson,
  Phys.\ Rev.\ Lett.\  {\bf 82}, 4971 (1999)
  [arXiv:hep-th/9803132].

\bibitem{Hsu:2004ri}
  S.~D.~H.~Hsu,
  Phys.\ Lett.\  B {\bf 594}, 13 (2004)
  [arXiv:hep-th/0403052].

\bibitem{Li:2004rb}
  M.~Li,
  Phys.\ Lett.\  B {\bf 603}, 1 (2004)
  [arXiv:hep-th/0403127].

\bibitem{Spergel:2006hy}
  D.~N.~Spergel {\it et al.}  [WMAP Collaboration],
  Astrophys.\ J.\ Suppl.\  {\bf 170}, 377 (2007)
  [arXiv:astro-ph/0603449].

\bibitem{Komatsu:2008hk}
  E.~Komatsu {\it et al.}  [WMAP Collaboration],
  arXiv:0803.0547 [astro-ph].

\bibitem{Riess:2006fw}
  A.~G.~Riess {\it et al.},
  Astrophys. J. 659, 98 (2007) [arXiv:astro-ph/0611572].

\bibitem{Feng:2004ad}
  B.~Feng, X.~L.~Wang and X.~M.~Zhang,
  Phys.\ Lett.\  B {\bf 607}, 35 (2005)
  [arXiv:astro-ph/0404224].

\bibitem{Zhao:2006qg}
  G.~B.~Zhao, J.~Q.~Xia, H.~Li, C.~Tao, J.~M.~Virey, Z.~H.~Zhu and X.~Zhang,
  Phys.\ Lett.\  B {\bf 648}, 8 (2007)
  [arXiv:astro-ph/0612728].

\bibitem{Zhao:2006bt}
  G.~B.~Zhao, J.~Q.~Xia, B.~Feng and X.~Zhang,
  Int.\ J.\ Mod.\ Phys.\  D {\bf 16}, 1229 (2007)
  [arXiv:astro-ph/0603621].

\bibitem{Wang:2006ts}
  Y.~Wang and P.~Mukherjee,
  Astrophys.\ J.\  {\bf 650}, 1 (2006)
  [arXiv:astro-ph/0604051].


\bibitem{Cai:2007gs}
  Y.~F.~Cai, M.~Z.~Li, J.~X.~Lu, Y.~S.~Piao, T.~T.~Qiu and X.~M.~Zhang,
  Phys.\ Lett.\  B {\bf 651}, 1 (2007)
  [arXiv:hep-th/0701016].

\bibitem{Xia:2007km}
  J.~Q.~Xia, Y.~F.~Cai, T.~T.~Qiu, G.~B.~Zhao and X.~Zhang,
  Int.\ J.\ Mod.\ Phys.\  D {\bf 17}, 1229 (2008)
  [arXiv:astro-ph/0703202].

\bibitem{Li:2005fm}
  M.~Z.~Li, B.~Feng and X.~M.~Zhang,
  JCAP {\bf 0512}, 002 (2005)
  [arXiv:hep-ph/0503268].

\bibitem{ArmendarizPicon:2004pm}
  C.~Armendariz-Picon,
  JCAP {\bf 0407}, 007 (2004)
  [arXiv:astro-ph/0405267];
  H.~Wei and R.~G.~Cai,
  Phys.\ Rev.\  D {\bf 73}, 083002 (2006)
  [arXiv:astro-ph/0603052].

\bibitem{Koivisto:2008xf}
  T.~S.~Koivisto and D.~F.~Mota,
  JCAP {\bf 0808}, 021 (2008)
  [arXiv:0805.4229 [astro-ph]].

\bibitem{Cai:2005ie}
  R.~G.~Cai, H.~S.~Zhang and A.~Wang,
  Commun.\ Theor.\ Phys.\  {\bf 44}, 948 (2005)
  [arXiv:hep-th/0505186];
  P.~S.~Apostolopoulos and N.~Tetradis,
  Phys.\ Rev.\  D {\bf 74}, 064021 (2006)
  [arXiv:hep-th/0604014];
  K.~Bamba, C.~Q.~Geng, S.~Nojiri and S.~D.~Odintsov,
  Phys. Rev. D 79, 083014 (2009)
  arXiv:0810.4296 [hep-th].

\bibitem{Aref'eva:2005fu}
  I.~Y.~Aref'eva, A.~S.~Koshelev and S.~Y.~Vernov,
  Phys.\ Rev.\  D {\bf 72}, 064017 (2005)
  [arXiv:astro-ph/0507067];
  S.~Y.~Vernov,
  arXiv:astro-ph/0612487;
  A.~S.~Koshelev,
  JHEP {\bf 0704}, 029 (2007)
  [arXiv:hep-th/0701103];
  Y.~F.~Cai and W.~Xue,
  Phys. Lett. B 680, 395 (2009)
  arXiv:0809.4134 [hep-th].

\bibitem{Guo:2004fq}
  Z.~K.~Guo, Y.~S.~Piao, X.~M.~Zhang and Y.~Z.~Zhang,
  Phys.\ Lett.\ {\bf B608}, 177 (2005)
  [arXiv:astro-ph/0410654].

\bibitem{Quintom_tf}
  X.~F.~Zhang, H.~Li, Y.~S.~Piao and X.~M.~Zhang,
  Mod.\ Phys.\ Lett.\  A {\bf 21}, 231 (2006)
  [arXiv:astro-ph/0501652];
  Z.~K.~Guo, Y.~S.~Piao, X.~Zhang and Y.~Z.~Zhang,
  Phys.\ Rev.\  D {\bf 74}, 127304 (2006)
  [arXiv:astro-ph/0608165].


\bibitem{Cai:2006dm}
  Y.~F.~Cai, H.~Li, Y.~S.~Piao and X.~M.~Zhang,
  Phys.\ Lett.\  B {\bf 646}, 141 (2007)
  [arXiv:gr-qc/0609039].

\bibitem{Quintom_1}
  B.~Feng, M.~Li, Y.~S.~Piao and X.~Zhang,
  Phys.\ Lett.\  B {\bf 634}, 101 (2006)
  [arXiv:astro-ph/0407432];
  H.~Wei and R.~G.~Cai,
  Phys.\ Rev.\  D {\bf 72}, 123507 (2005)
  [arXiv:astro-ph/0509328];
  X.~Zhang,
  Phys.\ Rev.\  D {\bf 74}, 103505 (2006)
  [arXiv:astro-ph/0609699].

\bibitem{Cai:2007qw}
  Y.~F.~Cai, T.~Qiu, Y.~S.~Piao, M.~Li and X.~Zhang,
  JHEP {\bf 0710}, 071 (2007)
  [arXiv:0704.1090 [gr-qc]].

\bibitem{Cai:2007zv}
  quintom is found to be able to give a bouncing solution and
  its perturbations combine aspects of singular and nonsingular
  bounce models, see for example:
  Y.~F.~Cai, T.~Qiu, R.~Brandenberger, Y.~S.~Piao and X.~Zhang,
  JCAP {\bf 0803}, 013 (2008)
  [arXiv:0711.2187 [hep-th]];
  Y.~F.~Cai, T.~t.~Qiu, R.~Brandenberger and X.~m.~Zhang,
  Phys. Rev. D 80, 023511 (2009)
  arXiv:0810.4677 [hep-th].

\bibitem{Cai:2008ed}
  Y.~F.~Cai, T.~T.~Qiu, J.~Q.~Xia and X.~Zhang,
  Phys. Rev. D 79, 021303 (2009)
  arXiv:0808.0819 [astro-ph];
  Y.~F.~Cai and X.~Zhang,
  JCAP 0906, 003 (2009)
  arXiv:0808.2551 [astro-ph].

\bibitem{Quintom_others}
  W.~Zhao, and Y. Zhang,
  Phys.\ Rev.\  D {\bf 73}, 123509 (2006)
  [arXiv:astro-ph/0604460];
  H.~Mohseni Sadjadi and M.~Alimohammadi,
  Phys.\ Rev.\  D {\bf 74}, 043506 (2006)
  [arXiv:gr-qc/0605143];
  E.~O.~Kahya and V.~K.~Onemli,
  Phys.\ Rev.\  D {\bf 76}, 043512 (2007)
  [arXiv:gr-qc/0612026];
  Y.~F.~Cai and Y.~S.~Piao,
  Phys.\ Lett.\  B {\bf 657}, 1 (2007)
  [arXiv:gr-qc/0701114].
  R.~Lazkoz, G.~Leon and I.~Quiros,
  Phys.\ Lett.\  B {\bf 649}, 103 (2007)
  [arXiv:astro-ph/0701353];
  H.~Zhang and Z.~H.~Zhu,
  Mod. Phys. Lett. A 24, 541 (2009)
  arXiv:0704.3121 [astro-ph];
  T.~Qiu, Y.~F.~Cai and X.~M.~Zhang,
  Mod. Phys. Lett. A 23, 2787 (2008)
  arXiv:0710.0115 [gr-qc];
  M.~R.~Setare, J.~Sadeghi and A.~Banijamali,
  Phys.\ Lett.\  B {\bf 669}, 9 (2008)
  [arXiv:0807.0077 [hep-th]].

\bibitem{Xiong:2007cn}
  H.~H.~Xiong, T.~Qiu, Y.~F.~Cai and X.~Zhang,
  Mod. Phys. Lett. A 24, 1237 (2009)
  arXiv:0711.4469 [hep-th];
  H.~H.~Xiong, Y.~F.~Cai, T.~Qiu, Y.~S.~Piao and X.~Zhang,
  Phys.\ Lett.\  B {\bf 666}, 212 (2008)
  [arXiv:0805.0413 [astro-ph]];
  S.~Li, Y.~F.~Cai and Y.~S.~Piao,
  Phys. Lett. B 671, 423 (2009)
  arXiv:0806.2363 [hep-ph];
  S.~Zhang and B.~Chen,
  Phys.\ Lett.\  B {\bf 669}, 4 (2008)
  [arXiv:0806.4435 [hep-ph]].

\bibitem{Elizalde:2004mq}
  E.~Elizalde, S.~Nojiri and S.~D.~Odintsov,
  Phys.\ Rev.\  D {\bf 70}, 043539 (2004)
  [arXiv:hep-th/0405034].

\bibitem{Koivisto:2007bp}
  T.~Koivisto and D.~F.~Mota,
  Astrophys.\ J.\  {\bf 679}, 1 (2008)
  [arXiv:0707.0279 [astro-ph]].

\bibitem{Cai:2008gk}
  Y.~F.~Cai and J.~Wang,
  Class.\ Quant.\ Grav.\  {\bf 25}, 165014 (2008)
  [arXiv:0806.3890 [hep-th]].

\bibitem{Veneziano:1991ek}
  G.~Veneziano,
  Phys.\ Lett.\  B {\bf 265}, 287 (1991).

\bibitem{Lidsey:1999mc}
  J.~E.~Lidsey, D.~Wands and E.~J.~Copeland,
  Phys.\ Rept.\  {\bf 337}, 343 (2000)
  [arXiv:hep-th/9909061].

\bibitem{Chimento:2003qy}
  L.~P.~Chimento and R.~Lazkoz,
  Phys.\ Rev.\ Lett.\  {\bf 91}, 211301 (2003)
  [arXiv:gr-qc/0307111].

\bibitem{Dabrowski:2003jm}
  M.~P.~Dabrowski, T.~Stachowiak and M.~Szydlowski,
  Phys.\ Rev.\  D {\bf 68}, 103519 (2003)
  [arXiv:hep-th/0307128].

\bibitem{Chimento:2005xa}
  L.~P.~Chimento and D.~Pavon,
  Phys.\ Rev.\  D {\bf 73}, 063511 (2006)
  [arXiv:gr-qc/0505096].

\bibitem{Dabrowski:2006dd}
  M.~P.~Dabrowski, C.~Kiefer and B.~Sandhofer,
  Phys.\ Rev.\  D {\bf 74}, 044022 (2006)
  [arXiv:hep-th/0605229].

\bibitem{Chiba:1998tc}
  T.~Chiba and T.~Nakamura,
  Prog.\ Theor.\ Phys.\  {\bf 100}, 1077 (1998)
  [arXiv:astro-ph/9808022].

\bibitem{Sahni:2002fz}
  V.~Sahni, T.~D.~Saini, A.~A.~Starobinsky and U.~Alam,
  JETP Lett.\  {\bf 77}, 201 (2003)
  [Pisma Zh.\ Eksp.\ Teor.\ Fiz.\  {\bf 77}, 249 (2003)]
  [arXiv:astro-ph/0201498].

\bibitem{Zimdahl:2003wg}
  W.~Zimdahl and D.~Pavon,
  Gen.\ Rel.\ Grav.\  {\bf 36}, 1483 (2004)
  [arXiv:gr-qc/0311067].

\bibitem{Alam:2003sc}
  U.~Alam, V.~Sahni, T.~D.~Saini and A.~A.~Starobinsky,
  Mon.\ Not.\ Roy.\ Astron.\ Soc.\  {\bf 344}, 1057 (2003)
  [arXiv:astro-ph/0303009].

\bibitem{Gorini:2002kf}
  V.~Gorini, A.~Kamenshchik and U.~Moschella,
  Phys.\ Rev.\  D {\bf 67}, 063509 (2003)
  [arXiv:astro-ph/0209395].

\bibitem{Zhang:2005rj}
  X.~Zhang,
  Phys.\ Lett.\  B {\bf 611}, 1 (2005)
  [arXiv:astro-ph/0503075].

\bibitem{Zhang:2005yz}
  X.~Zhang,
  Int.\ J.\ Mod.\ Phys.\  D {\bf 14}, 1597 (2005)
  [arXiv:astro-ph/0504586].

\bibitem{Wu:2005apa}
  P.~X.~Wu and H.~W.~Yu,
  Int.\ J.\ Mod.\ Phys.\  D {\bf 14}, 1873 (2005)
  [arXiv:gr-qc/0509036].

\bibitem{Zhang:2004gc}
  X.~Zhang, F.~Q.~Wu and J.~Zhang,
  JCAP {\bf 0601}, 003 (2006)
  [arXiv:astro-ph/0411221].

\bibitem{Setare:2006xu}
  M.~R.~Setare, J.~Zhang and X.~Zhang,
  JCAP {\bf 0703}, 007 (2007)
  [arXiv:gr-qc/0611084].

\bibitem{Yi:2007gu}
  Z.~L.~Yi and T.~J.~Zhang,
  Phys.\ Rev.\  D {\bf 75}, 083515 (2007)
  [arXiv:astro-ph/0703630].


\bibitem{Zhao:2007jn}
  W.~Zhao,
  arXiv:0711.2319 [gr-qc].

\bibitem{Panotopoulos:2007zn}
  G.~Panotopoulos,
  Nucl.\ Phys.\  B {\bf 796}, 66 (2008)
  [arXiv:0712.1177 [astro-ph]].

\bibitem{Huang:2008gs}
  Z.~G.~Huang, X.~M.~Song, H.~Q.~Lu and W.~Fang,
  Astrophys.\ Space Sci.\  {\bf 315}, 175 (2008)
  [arXiv:0802.2320 [hep-th]].

\bibitem{Shao:2008zz}
  Y.~Shao and H.~Zhong,
  Mod.\ Phys.\ Lett.\  A {\bf 23}, 879 (2008).

\bibitem{Feng:2008rs}
  C.~J.~Feng,
  arXiv:0809.2502 [hep-th];
  C.~J.~Feng,
  arXiv:0810.2594 [hep-th].

\bibitem{Chimento:2002gb}
  L.~P.~Chimento,
  Phys.\ Rev.\  D {\bf 65}, 063517 (2002).

\bibitem{Aguirregabiria:2003uh}
  J.~M.~Aguirregabiria, L.~P.~Chimento, A.~S.~Jakubi and R.~Lazkoz,
  Phys.\ Rev.\  D {\bf 67}, 083518 (2003)
  [arXiv:gr-qc/0303010].

\bibitem{Aguirregabiria:2004te}
  J.~M.~Aguirregabiria, L.~P.~Chimento and R.~Lazkoz,
  Phys.\ Rev.\  D {\bf 70}, 023509 (2004)
  [arXiv:astro-ph/0403157].

\bibitem{Chimento4} L. P. Chimento, R. Lazkoz, Class. Quant. Grav. {\bf 23}, 3195
(2006).

\bibitem{Chimento:2006gk}
  L.~P.~Chimento and W.~Zimdahl,
  Int.\ J.\ Mod.\ Phys.\  D {\bf 17}, 2229 (2008)
  [arXiv:gr-qc/0609104].

\bibitem{Chimento:2007fx}
  L.~P.~Chimento, F.~P.~Devecchi, M.~Forte and G.~M.~Kremer,
  Class.\ Quant.\ Grav.\  {\bf 25}, 085007 (2008)
  [arXiv:0707.4455 [gr-qc]].

\bibitem{ArmendarizPicon:2003qk}
  C.~Armendariz-Picon and P.~B.~Greene,
  Gen.\ Rel.\ Grav.\  {\bf 35}, 1637 (2003)
  [arXiv:hep-th/0301129].

\bibitem{Vakili:2005ya}
  B.~Vakili, S.~Jalalzadeh and H.~R.~Sepangi,
  JCAP {\bf 0505}, 006 (2005)
  [arXiv:gr-qc/0502076].

\bibitem{Ribas:2005vr}
  M.~O.~Ribas, F.~P.~Devecchi and G.~M.~Kremer,
  Phys.\ Rev.\  D {\bf 72}, 123502 (2005)
  [arXiv:gr-qc/0511099].

\bibitem{Weinberg}
  S. Weinberg, \textit{Gravitation and Cosmology}, Cambridge University Press (1972).

\bibitem{BirrellDavies}
  N. Birrell and P. Davies, \textit{Quantum Fields in Curved Space}, Cambridge University Press (1982).

\bibitem{GSW}
  M. Green, J.  Schwarz, E.   Witten, \textit{Superstring Theory Vol. 2}, Chapter 12, Cambridge University Press (1987).



\end{thebibliography}
\end{document}